\documentstyle[12pt,aasms4]{article}

\begin{document}

\noindent
{\it Conference Highlights}

\begin{center}


\title{\large \bf International Workshop on Redshift Mechanisms in 
Astrophysics and Cosmology\footnote{Conference was held in Clonakilty-Cork, 
Ireland in May 15-18, 2006.}} 

\end{center}

\medskip


An extraordinary event took place recently in Ireland. A group of independent 
and professional researchers met to discuss an old heterodox topic with 
important consequences in astrophysics and, especially, in cosmology: possible 
causes of the redshifts in the spectra of astrophysical objects other than a 
Doppler or expanding universe mechanism. Many decades of work have been 
devoted to this kind of research, most of it forgotten by the greater part 
of the astrophysical community nowadays. But the question is still open, the 
debate is still alive, as was shown by the participants in the present 
Workshop. There is no smoke without fire, and the existence of many facts 
and theories on alternative origins of redshifts may point to some new
pathways in physics that deserve further attention. This was precisely the 
aim of this meeting.

Observational facts were presented by Tom Van Flandern (Meta Research, Sequim, 
WA, U.S.A.), Chris Fulton (James Cook University, Australia; in collaboration 
with Halton Arp, MPIA, Germany), William M. Napier (University of Cardiff, 
U. K.) and myself (Mart\'\i n L\'opez-Corredoira, IAC, Tenerife, Spain, in 
collaboration with Carlos M. Guti\'errez of the same institute).  

Tom Van Flandern sketched ``A User's Guide to the Universe'' to 
illustrate the fact that the problems with the Big Bang are numerous and 
basic, starting right from the ideas that redshift implies expansion or 
that the microwave radiation is from the ``background''. In particular, 
the supernova data, once thought to provide the best evidence for expansion, 
now appears to leave no room for expansion after correction for ``Malmquist 
bias''---the tendency to find only the brightest members of any class of 
objects at the greatest distances. And Eddington already showed in 1923 
that the natural ``temperature of space'' was 3 K because of radiation from 
distant star-and-galaxy light. There was a lively discussion with further 
presentations on the evidence for expansion of the universe, in which 
Bill Napier defended the idea that expansion is evident with the current 
proofs, while Tom Van Flandern and I adopted a more skeptical attitude. 

My presentation (L\'opez-Corredoira) was about the statistical correlations 
of high-redshift QSOs/galaxies and nearby galaxies, and an excess of QSOs 
along the minor axes of these galaxies. Gravitational lensing would seem 
insufficient to explain them. Some of these cases may be just fortuitous 
associations in which background objects are close in the sky to a foreground 
galaxy, although the statistical mean correlations remain to be explained and 
some lone objects, which can even show bridges connecting four objects with 
very different redshifts, have very small probabilities of being a projection 
of background objects. 

Bill Napier talked about the current situation of the Tifft and Karlsson 
effect. A galactocentric periodicity of 37.5 km/s has been claimed to exist in 
the redshift distribution of nearby spiral galaxies. This phenomenon was 
further investigated using both the original dataset of 97 galaxies, and 
a further two datasets, namely 42 galaxies with high-precision redshift 
measures, and ~102 galaxies within 500 km/s. The galaxies in the latter 
set have less precisely measured redshifts but have photometrically measured 
distances. The periodicity is seen in all the datasets examined. 
Using the distance information, there is some evidence that the periodicity 
is not strictly with respect to a fixed galactocentric velocity vector, 
but is referred to a vector which varies with increasing distance. With 
respect to the centre of the Galaxy, the Hubble constant is found to be 
61.1$\pm $5.0 km/s with intercept -4.7$\pm $18.2 km/s, 
with a remarkably small dispersion. Earlier claims of a 72 km/s periodicity in 
the redshifts of the Virgo cluster, and of a 0.089 periodicity in 
$\log_{10}(1+z)$ for QSOs close to active spiral galaxies, were also 
reviewed.

With reference to QSO redshift periodicities, Chris Fulton discussed his 
custom computer program to detect quasar families in the entire 2dF deep 
survey area. The program uses the positions, redshifts, and magnitudes of 
all galaxies and quasars, applies various empirically derived constraints 
to these data, and produces a catalogue of galaxies for which at least four 
or five quasars pass all applied constraints and are consequently identified 
as members of a family with real physical associations.

It is normally objected that the observed anomalies are not sufficient to 
support the existence of non-cosmological/Doppler redshifts because there 
is no theory that can explain them. Yet this is not true. In the recent 
workshop, a few hypotheses were presented among the broad range of theories 
which are available in the literature. These are due to Henrik Broberg, 
Amitabha Ghosh (Indian Institute of Technology, Kanpur, India), 
Tom Van Flandern (Meta Research, Sequim, WA, U.S.A.), David Roscoe 
(Sheffield University, U. K.), Jacques Moret-Bailly (University of 
Bourgogne, Dijon). In addition, Sisir Roy and collaborators 
(Malabika Roy and Menas Kafatos, all of George Mason University) sent 
a contribution via e-mail since they were unable to attend.

Some theories are based on heterodox ideas about gravitation. Henrik Broberg, 
for example, pointed out that a comparison between a quantizing scheme 
inherent in the Schwarzschild metric and observed redshifts from quasars 
shows that the contracted distances in the gravitational field are quantized 
in terms of gravitational radii of the gravitating objects responsible for the 
field, while non-contracted distances are not so quantized. This observation, 
in Broberg's view, might be as important for modern quantum physics as the 
Michelson-Morley Experiment was for the introduction of Special Relativity. 
Quantization, therefore, appears in systems with significant relativistic 
contraction (gravitational or Lorentz-contraction), such as de Broglie 
waves or, in the extreme case, photon energy packets at the speed of light. 
In generalised terms, this would mean that quantum physics has its origin in 
relativistically contracted fields, while the gravitational field can be 
scaled to particle dimensions, with a surface energy density constant ($A$) 
as the invariant parameter in the process. This might, for example, lead 
to the possibility of a topological transformation from the gravitational 
force to the strong nuclear force.

Amitabha Ghosh, meanwhile, proposed a phenomenological model of dynamic 
gravitational interaction between two objects that depends not only on 
their masses and the distance between them, but also on the relative 
velocity and acceleration between the two. These velocity- and 
acceleration-dependent interactions are termed ``inertial induction''. 
The velocity-dependent force leads to a cosmic drag on all objects moving 
with respect to the mean rest-frame of a universe treated as infinite and 
quasi-static. This cosmic drag results in the cosmological redshift of 
light coming from distant galaxies without any universal expansion. This 
force model leads to the exact equivalence of gravitational and inertial 
masses and explains many not well understood observed celestial and 
astronomical phenomena.

Tom Van Flandern demonstrated how a friction between gravitons and photons 
might be responsible for the redshift ($z$) because it provides an energy 
loss mechanism not subject to the usual problem of making distant galaxy 
images fuzzy, and because it varies with $(1+z)^{-2}$ 
which is closer to the observed rate of decrease of surface brightness 
than any Big Bang model. He also showed that if we adopt premises from 
deductive logic that need no miracles instead of interpretations arrived 
at through inductive guesswork from observations, we develop a more 
plausible way to understand and describe the universe than is 
available from the miracle-based Big Bang.  The ``principles of physics'' 
imply properties of dimensions and forces consistent with classical concepts. 
Perhaps the single, most important difference from current views, notes 
Van Flandern, is that the universe has no speed limit-consistent 
with extensive experimental evidence that the propagation speed of 
gravitational force is many orders of magnitude faster (in forward time) 
than the speed of light.

Instead of new ideas on gravitation, David Roscoe sought to cast new 
light on classical electrodynamics. After abstracting from certain 
well-known symmetries of classical electrodynamics and showing that 
these symmetries are alone sufficient to recover the classical theory 
exactly and unadorned, he found that the classical electromagnetic field is 
then "irreducibly" associated with a massive vector field. This latter field 
can only be interpreted as a classical representation of a massive photon. 
That is, it was demonstrated that existing classical theory is ``already'' 
compatible with the notion of a massive photon. He was then able to show that, 
given a certain natural assumption about how we should calculate the 
trajectories of this massive photon, the ``quantized redshift phenomenology'' 
can be understood in a simple and natural way.

Other new ideas rested not on new physics or new reinterpretations of old 
theories, but on effects predicted at present with standard physics. 
Jacques Moret-Bailly said that the Karlsson periodicities observed 
mainly in the spectra of the quasars could be understood using the 
CREIL effect (Coherent Raman Effect on Time-Incoherent Light): neutral 
atomic hydrogen is pumped by Lyman alpha absorption to the $2p$ state, 
producing a redshift which renews the intensity at the pumping frequency 
if the initial intensity is sufficient. An absorbed line almost stops the 
redshift, so that the Lyman beta and gamma are strongly absorbed; then the 
redshift restarts. It now appears that a Lyman absorption in cold molecular 
hydrogen can produce the Tifft-Napier periodicity of 72 km/s.

Another possible way-say Sisir Roy et al.--to explain the frequency 
shift of the spectral lines involves induced correlations in a random, 
inhomogeneous medium, as discovered by Emil Wolf. The shift mimics the 
Doppler mechanism even if there is no relative motion between the observer 
and the source. Sisir Roy and his collaborators have formulated a Dynamical 
Multiple Scattering (DMS) Theory based on Wolf's idea to account for the 
frequency shift in light coming from distant sources, such as quasars. 
Recent observations of molecular gases around quasars with high redshifts 
suggest that this medium is suitable to produce DMS. 

I feel this workshop was extraordinary in two important respects: in the 
challenge posed by the topic chosen; and in the relaxed way in which 
participants could exchange ideas, discuss, debate, and think creatively 
in an atmosphere of true passion for science. It is reminiscent of the 
golden era of physics when a few people would meet for hours and hours 
in friendly discussion of new ideas; something far removed from the 
typical scientific meetings organized nowadays in which hundreds 
of participants have a few minutes each to rush through expositions 
of boring work, often without any novelty (much less any challenge 
to accepted ideas), or they are led by few orthodox organizers to 
secure support for their own credos. Was it a dream, or perhaps an 
illusion inspired by nostalgia for times when the physics was not as 
dead as it is now? Certainly too beautiful to be true. It would be 
a bonus if some of the discussed ideas in the workshop were shown 
to be correct. This would mean that it is still possible to do first 
rank physics, rather than the usual second or third class science. 
Unfortunately, whether these ideas are correct or not, the scientific 
community is probably not yet ready to change its mind and abandon 
its dogma of Doppler/expansion redshifts, so we cannot know at 
present how much truth they contain. We will have to wait for the future, 
for Nature's answer. We should also bear in mind that Nature does not 
care about our illusions; the truth is independent of our nostalgias.

\medskip


\noindent
{\it Mart\'\i n L\'opez-Corredoira} 

\noindent
{\it Instituto de Astrof\'\i sica de Canarias

C/.V\'\i a L\'actea, s/n

E-38200 La Laguna (Tenerife, Spain) 

E-mail: martinlc@iac.es}

\end{document}